\documentclass[prd,twocolumn,showpacs,preprintnumbers,amsmath,nofootinbib,amssymb,floatfix]{revtex4-1}
\usepackage{graphicx}
\usepackage{epsfig,float}
\usepackage[sort&compress]{natbib}
\usepackage{subfigure}
\usepackage{amsmath}
\usepackage{amsfonts}
\usepackage{cancel}
\usepackage{amssymb}
\usepackage{hyperref}
\usepackage{color}
\usepackage{url}
\begin{document}
\arraycolsep1.5pt
\newcommand{\Ima}{\textrm{Im}}
\newcommand{\Rea}{\textrm{Re}}
\newcommand{\mev}{\textrm{ MeV}}
\newcommand{\gev}{\textrm{ GeV}}
\newcommand{\dtres}{d^{\hspace{0.1mm} 3}\hspace{-0.5mm}}
\newcommand{\rts}{ \sqrt s}
\newcommand{\non}{\nonumber \\[2mm]}
\newcommand{\eps}{\epsilon}
\newcommand{\half}{\frac{1}{2}}
\newcommand{\thalf}{\textstyle \frac{1}{2}}
\newcommand{\Nmass}{M_{N}} 
\newcommand{\delmass}{M_{\Delta}} 
\newcommand{\pimass}{\mu}  
\newcommand{\rhomass}{m_\rho} 
\newcommand{\piNN}{f}      
\newcommand{\rhocoup}{g_\rho} 
\newcommand{\fpi}{f_\pi} 
\newcommand{\f}{f} 
\newcommand{\nucfld}{\psi_N} 
\newcommand{\delfld}{\psi_\Delta} 
\newcommand{\fpiNN}{f_{\pi N N}} 
\newcommand{\fpiND}{f_{\pi N \Delta}} 
\newcommand{\GMquark}{G^M_{(q)}} 
\newcommand{\vecpi}{\vec \pi}
\newcommand{\vectau}{\vec \tau}
\newcommand{\vecrho}{\vec \rho}
\newcommand{\delmu}{\partial_\mu}
\newcommand{\delMu}{\partial^\mu}
\newcommand{\nn}{\nonumber}
\newcommand{\bi}{\bibitem}
\newcommand{\vs}{\vspace{-0.20cm}}
\newcommand{\be}{\begin{equation}}
\newcommand{\ee}{\end{equation}}
\newcommand{\ba}{\begin{eqnarray}}
\newcommand{\ea}{\end{eqnarray}}
\newcommand{\ropi}{$\rho \rightarrow \pi^{0} \pi^{0}
\gamma$ }
\newcommand{\roeta}{$\rho \rightarrow \pi^{0} \eta
\gamma$ }
\newcommand{\omepi}{$\omega \rightarrow \pi^{0} \pi^{0}
\gamma$ }
\newcommand{\omeeta}{$\omega \rightarrow \pi^{0} \eta
\gamma$ }
\newcommand{\ul}{\underline}
\newcommand{\del}{\partial}
\newcommand{\rth}{\frac{1}{\sqrt{3}}}
\newcommand{\rsix}{\frac{1}{\sqrt{6}}}
\newcommand{\sq}{\sqrt}
\newcommand{\fr}{\frac}
\newcommand{\pr}{^\prime}
\newcommand{\ov}{\overline}
\newcommand{\Gm}{\Gamma}
\newcommand{\rw}{\rightarrow}
\newcommand{\rgl}{\rangle}
\newcommand{\De}{\Delta}
\newcommand{\Dp}{\Delta^+}
\newcommand{\Dm}{\Delta^-}
\newcommand{\Dz}{\Delta^0}
\newcommand{\Dpp}{\Delta^{++}}
\newcommand{\Sg}{\Sigma^*}
\newcommand{\Sp}{\Sigma^{*+}}
\newcommand{\Sm}{\Sigma^{*-}}
\newcommand{\Sz}{\Sigma^{*0}}
\newcommand{\X}{\Xi^*}
\newcommand{\Xm}{\Xi^{*-}}
\newcommand{\Xz}{\Xi^{*0}}
\newcommand{\Om}{\Omega}
\newcommand{\Omm}{\Omega^-}
\newcommand{\kp}{K^+}
\newcommand{\kz}{K^0}
\newcommand{\pip}{\pi^+}
\newcommand{\pim}{\pi^-}
\newcommand{\piz}{\pi^0}
\newcommand{\et}{\eta}
\newcommand{\kb}{\ov K}
\newcommand{\km}{K^-}
\newcommand{\kbz}{\ov K^0}
\newcommand{\ksb}{\ov {K^*}}

\newcommand{\re}{\text{Re}}
\newcommand{\im}{\text{Im}}

\title{ $\chi_{c0}(1P)$ decay into $\bar \Sigma~ \Sigma \pi$ in search of an  $I=1$, $1/2^-$ baryon state around $\bar K N$ threshold}

\author{En Wang} \email{wangen@zzu.edu.cn}
\affiliation{Department of Physics, Zhengzhou University, Zhengzhou, Henan 450001, China
}
\author{Ju-Jun Xie} \email{xiejujun@impcas.ac.cn}
\affiliation{Institute of Modern Physics, Chinese Academy of
Sciences, Lanzhou 730000, China}
\affiliation{State Key Laboratory of Theoretical
Physics, Institute of Theoretical Physics, Chinese Academy of
Sciences, Beijing 100190, China}
\author{Eulogio Oset} \email{oset@ific.uv.es}
\affiliation{Departamento de F\'{\i}sica Te\'orica and IFIC, Centro Mixto Universidad de Valencia-CSIC Institutos de Investigaci\'on de Paterna, Aptdo.22085, 46071 Valencia, Spain}
\affiliation{Institute of Modern Physics, Chinese Academy of
Sciences, Lanzhou 730000, China}
\date{\today}

\begin{abstract}
We present the theoretical study of the process $\chi_{c0}(1P)\to\bar\Sigma\Sigma\pi $ decay, by  taking into account the $\pi\Sigma$ and $\pi\bar\Sigma$ final state interactions of the final meson-baryon pair based on the chiral unitary approach. We show that the process filters the isospin $I=1$ in the $\pi\Sigma$ channel and offers a reaction to test the existence of an $I=1$ state with strangeness $S=-1$ and spin-parity $J^p=1/2^-$ around the $\bar KN$ threshold predicted by some theories and supported by some experiments.

\end{abstract}


\maketitle

\section{Introduction}
\label{Intro} 

The extraction of baryon resonances from experimental data is one of the important aims in Hadron Physics and much progress has been done in the latest years \cite{Klempt:2009pi,Aznauryan:2011ub,Crede:2013sze}. The traditional tools to learn about these resonances have been the use of pion beams \cite{Manley:1984jz,Arndt:2006bf}, photon beams \cite{Aznauryan:2011ub,Anisovich:2011fc} and also kaon beams \cite{Prakhov:2004an}. The advent of new facilities as BES, CDF, LHCb is also contributing to enlarge the list of baryon resonances 
\cite{Ablikim:2004ug,Aaltonen:2011sf,Aaij:2012da,Aaij:2015tga,pdg}. On the other hand, the 
theoretical work goes parallel and many predictions are made. The quark models jumped earlier in this arena \cite{Isgur:1978wd,Capstick:2000qj,Valcarce:1995dm}, but effective theories have also contributed their share \cite{cola,Wu:2010jy,Wu:2010rv,Romanets:2012hm}. The quark models seem to over predict the number of baryon states, giving rise to the problem of the missing resonances. The effective theories give rise to some dynamically generated states as a consequence of the interaction of two hadrons, which fit some of the existing states, and also predict new states, most of them in the heavy sectors of charm and beauty. Some of the predictions of these effective theories have been confirmed experimentally.  One of the clear cases is the existence of two $\Lambda(1405)$ states, which were first reported in Ref.~\cite{ollerulf}, discussed in detail in Ref.~\cite{cola} and later on confirmed in
 all theories using the chiral unitary approach  
\cite{Kaiser:1995eg,Oset:1997it,Hyodo:2002pk,GarciaRecio:2002td,GarciaRecio:2005hy,Borasoy:2005ie,Oller:2006jw,
Borasoy:2006sr,hyodonew,kanchan,hyodorev,ollerguo,maimeissner}. The experimental confirmation came from the work of Ref.~\cite{Prakhov:2004an} and the analysis of Ref.~\cite{magasramos}, but other experiments have come to confirm it too (see the introduction Ref. \cite{luispho1} for details)\footnote{The PDG will introduce officially the two $\Lambda(1405)$ states in the next Edition of the Book}. Alternative, pictures for some $N^*$ states and the $\Lambda(1405)$ involving pentaquarks have also been invoked~\cite{Liu:2005pm,Zou:2006uh}. Although not identical to the molecular representation, the need for more than standard three quarks is also deduced in those works.
  Another one of these successful predictions is the existence of a $D$ state with spin zero at $2600$ MeV and a width of about 100 MeV from the interaction of $\rho$ ($\omega$) and $D^*$ \cite{hidekoraquel}. This state is also in agreement with the $D(2600)$, with a similar width, discovered after the theoretical work in Ref.~\cite{delAmoSanchez:2010vq}.
The list of  predicted states which has found experimental support have been found is long (see Ref.~\cite{miguexie} as an example). 
Although it is premature to judge, the recent narrow pentaquark reported by the LHCb collaboration \cite{Aaij:2015tga} could maybe correspond to the predictions made of a hidden charm state in Ref.~\cite{Wu:2010jy} (see \cite{penta}).

    With this favorable perspective, the purpose of this paper is to call the attention to a possible intriguing baryon state of $J^P=1/2^-$, strangeness $S=-1$ and isospin $I=1$ around 1430 MeV, predicted in theories using the chiral unitary approach. This state shows up in the work of Ref.~\cite{ollerulf} and becomes a pronounced cusp (corresponding to a virtual state) in Ref.~\cite{cola}. One should note that such borderline states are common and one of them, classified as a resonance in the PDG, is the $a_0(980)$, as found in Ref.~\cite{ramonet,  dany,Pelaez:2006nj,Pelaez:2003dy,Baru:2003qq,Branz:2008ha,Wolkanowski:2015lsa}. The existence of this state $I=1$ has also been claimed from a different perspective in Ref.~\cite{Wu:2009nw}.

   One of the experiments that has brought some light on this state is the photoproduction of the 
$\Lambda(1405)$ undertaken in Refs.~\cite{Niiyama:2008rt,Moriya:2012zz,Moriya:2013eb,Moriya:2013hwg} and analysed in Ref.~\cite{luispho2}. The analyses in the experimental papers and in the theoretical ones differ in the predictions with respect to this state, although the two approaches lead to $I=1$ states. We should note that the analysis of Ref.~\cite{luispho2} preserves unitarity in coupled channels, analyticity and all relevant properties of the scattering matrices, while some approximations are done in Refs.~ \cite{Moriya:2012zz,Moriya:2013eb,Moriya:2013hwg}. The result of  Ref.~\cite{luispho2} is that there is a state of $I=1$, around the $\bar K N$ threshold, similar to the $a_0(980)$, visible as a strong cusp and in agreement with the findings of Refs.~\cite{ollerulf,cola}. It is clear that the extraction of this state is very problematic in conventional reactions which mix $I=0$ and $I=1$, and make it difficult to disentangle the $I=1$ contribution which, however, is of great importance to understand why are there such large differences in the shapes of the mass distributions of $\pi^+ \Sigma^-$, $\pi^- \Sigma^+$, $\pi^0 \Sigma^0$. 

In view of these problems we propose here a completely different method, feasible in present experimental facilities. The reaction proposed is $\chi_{c0}(1P) \to \bar \Sigma \Sigma\pi $. The reaction  $\chi_{c0}(1P) \to \bar \Sigma  \Sigma$ has been measured at BESIII~\cite{Ablikim:2012ena} and CLEO 
 \cite{Naik:2008zz} and the branching ratios are of order of $10^{-3}$. On the other hand, by looking at the PDG we find that the branching ratio for the analogous reaction 
 $\chi_{c0}(1P) \to \bar p \pi p$ is about three times larger than  that of the $\chi_{c0}(1P) \to \bar p p$, without $\pi$ production. One is then talking about branching ratios of the order of $3\times 10^{-3}$, easily accessible at BESIII. Given the quantum numbers of the $\chi_{c0}(1P)$, $I^G(J^{PC})=0^+(0^{++})$ and the fact that the $\chi_{c0}(1P)$ is a $c \bar c$ state, blind to SU(3), hence behaving like an SU(3) singlet, since the $\bar \Sigma$ has isospin $I=1$, the $\pi \Sigma$ state must have also $I=1$, to combine to the $I=0$ of the $\chi_{c0}(1P)$. This is a good filter of isospin that guarantees that the $\pi \Sigma$ will be in $I=1$. The $I=1$ state shows up more strongly in the $\pi \Sigma \to \pi \Sigma$ amplitude than in $\pi\Sigma\to\pi\Lambda$~\cite{luispho2}, so the $\pi\Sigma$ final state is the ideal channel within the approach used here.
 
The idea followed here to filter $I=1$ has its precedent in the studies of $J/\psi$ decay into $p\bar{p}$ and a meson~\cite{LI:1999du,Zou:2000wg}. Indeed since $J/\psi$ has $I=0$ and $\bar{p}$ $I=1/2$, the combination of $p$ and the meson will be in $I=1/2$. This idea was used in Ref.~\cite{Zou:1999wd} to study the $J/\psi\to \bar{p}N^*(1440)$, in Ref.~\cite{Liang:2002tk} to study the $J/\psi\to p\bar{p}\omega$ reaction and in Ref.~\cite{Liang:2004sd} to study the decays $J/\psi\to p\bar{p}\pi(\eta, \eta', \omega)$.
 
  We study the reaction and evaluate the $\pi \Sigma$ mass distribution and we find that indeed, the filter works well and a clear signal for this state, with practically no background from the $\pi \Sigma$ amplitude, is found.
  We then propose the implementation of this reaction which should settle this issue definitely and might result in the observation of a new baryon resonance in the light sector. 
  
  The present paper is organized as follows. In Sec.~\ref{sec:formalism}, we
shall discuss the formalism and the main ingredients of the model.
In Sec.~\ref{sec:results}, we will present our main results and
finally, the conclusions will be given in
Sec.~\ref{sec:conclusions}.
 
\section{Formalism}\label{sec:formalism}
In this section, we will describe the reaction mechanism for the process of $\chi_{c0}\rightarrow\bar \Sigma\Sigma\pi$. 
\subsection{The model of {\boldmath$\chi_{c0}\rightarrow\bar \Sigma\Sigma\pi$} }
\begin{figure}[!htb]
\centering
  \includegraphics[width=0.45\textwidth]{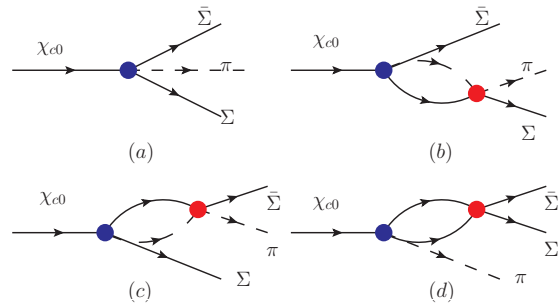}
\caption{(Color online) Diagrams for the $\chi_{c0}\rightarrow\bar \Sigma\Sigma\pi$ decay: (a) direct $\bar \Sigma\Sigma\pi$ vertex at tree level, (b) final state interaction of $\pi\Sigma$, (c) final state interaction of $\pi\bar\Sigma$, and (d) final state interaction of $\bar{\Sigma}\Sigma$.} 
  \label{fig:decaymodel}
\end{figure}

We will assume a contact interaction for $\chi_{c0}\rightarrow\bar \Sigma\Sigma\pi$ as a primary step [see Fig.~\ref{fig:decaymodel}(a)], and then we shall let the particles undergo final state interaction. This is the method to produce the dynamically generated resonances, in this case, the $I=1$ state, since they emerge as a consequence of the interaction of pairs of hadrons.

When considering the $\pi\Sigma$ final state [see Fig.~\ref{fig:decaymodel}(b)], one must take into account that in the first step one can produce other meson-baryon pairs that couple to the same $\pi\Sigma$ quantum numbers, then reaching the final $\pi\Sigma$ state through re-scattering. This forces us to see the possible meson baryon combinations in the first step. To this purpose, we must consider that the $\chi_{c0}$ is a SU(3) singlet, hence, since the $\bar\Sigma$ belongs to an SU(3) octet, then the $\pi\Sigma$ system will also be in an octet state. Since both the $\pi$ and $\Sigma$ belong to SU(3) octets, then we have the same situation as in the Yukawa coupling and we have two independent representation for $8(\pi)\otimes 8(\Sigma)$ going to $8^s$  and $8^a$. Technically, we can use an effective Lagrangian of the type
\begin{equation}
\mathcal{L} \equiv \tilde{D}\left\langle \bar{B}\left\lbrace\Phi,B\right\rbrace\right\rangle+ \tilde{F}\left\langle \bar{B}\left[ B, \Phi\right]\right\rangle,
\end{equation}
where the symbol $\left\langle~\right\rangle$ stands for the trace of SU(3) matrices, and the matrices corresponding to the octet of mesons and octet of baryons are the following,
\begin{equation}
\Phi = \left( 
\begin{array}{ccc}
\frac{1}{\sqrt{2}}\pi^0+\frac{1}{\sqrt{6}}\eta & \pi^+ & K^+ \\
\pi^- & -\frac{1}{\sqrt{2}}\pi^0+\frac{1}{\sqrt{6}}\eta & K^0 \\
K^- & \bar{K}^0 & -\frac{2}{\sqrt{6}}\eta
\end{array} \right)\,,
\end{equation}
\begin{equation}
B=\left(\begin{array}{ccc}
\frac{1}{\sqrt{2}}\Sigma^0+\frac{1}{\sqrt{6}}\Lambda & \Sigma^+ & p \\
\Sigma^- & -\frac{1}{\sqrt{2}}\Sigma^0+\frac{1}{\sqrt{6}}\Lambda & n \\
\Xi^- & \Xi^0 & -\frac{2}{\sqrt{6}}\Lambda
\end{array}\right)\,.
\end{equation} 

By looking at the SU(3) isoscalar factors in the PDG~\citep{pdg}, we find the weights of Table~\ref{tab:weight} in the isospin basis for the states. The sum of the isoscalar coefficients times $\tilde{D}$ and $\tilde{F}$ gives the weights $h_i$, which go into the primary production of each meson baryon channel.
\begin{table}
\caption{SU(3) isoscalar coefficients for the $\left\langle\bar{\Sigma}|MB\right\rangle$ matrix elements.}
\label{tab:weight}
\begin{tabular}{c|ccccc}
\hline \hline
$\bar{\Sigma}$ & $\bar{K}N$ & $\pi\Sigma$ & $\pi\Lambda$ & $\eta\Sigma$ & $K\Xi$ \\ \hline
$\tilde{D}$ & $-\sqrt{\frac{3}{10}}$ & 0 & $\sqrt{\frac{1}{5}}$ & $\sqrt{\frac{1}{5}}$ & $-\sqrt{\frac{3}{10}}$ \\
$\tilde{F}$ & $\sqrt{\frac{1}{6}}$ & $\sqrt{\frac{2}{3}}$ & 0 & 0 & $-\sqrt{\frac{1}{6}}$ \\
\hline
\end{tabular}
\end{table}

Next, we incorporate the final state interaction of these meson-baryon pairs, which is depicted in Fig.~\ref{fig:decaymodel}(b) and (c). The amplitude $\mathcal{M}(M_{\pi\Sigma},M_{\pi\bar\Sigma})$ for the transition can be written as, 
\begin{eqnarray}
\mathcal{M}(M_{\pi\Sigma},M_{\pi\bar\Sigma})&=& V_p\left(h_{\pi\Sigma}+\sum_{i}h_iG_i(M_{\pi\Sigma})\,t_{i,\pi\Sigma}(M_{\pi\Sigma}) \right. \, \nonumber \\
&& \left. + \sum_{i}h_iG_i(M_{\pi\bar\Sigma})\,t_{i,\pi\bar\Sigma}(M_{\pi\bar\Sigma})\right) \, \nonumber \\
&=&V_p\left( h_{\pi\Sigma} + T_{\pi\Sigma} + T_{\pi\bar\Sigma}\right) ,\label{eqn:fullamplitude}
\end{eqnarray}
where $V_p$ expresses the strength of an amplitude with $h=1$, and
 $G_i$ denotes the one-meson-one-baryon loop function, chosen in accordance with the model for the scattering matrix $t_{ij}$ that will be described in the next section. $M_{\pi\Sigma}$ and  $M_{\pi\bar\Sigma}$ are the invariant masses of the final states $\pi\Sigma$ and $\pi\bar\Sigma$, respectively, and $h_i$ stands for the weights of the transition $\chi_{c0}\to BP\bar \Sigma$ at tree level, which are given by,
\begin{gather}
h_{\bar KN}= -\sqrt{\frac{3}{10}}\tilde{D} + \sqrt{\frac{1}{6}}\tilde{F},\nonumber \\ 
h_{\pi \Sigma}=\sqrt{\frac{2}{3}}\tilde{F},~~
h_{\pi\Lambda}=\sqrt{\frac{1}{5}}\tilde{D},~~
h_{\eta\Sigma}=\sqrt{\frac{1}{5}}\tilde{D},\nonumber \\
h_{K\Xi}=-\sqrt{\frac{3}{10}}\tilde{D}-\sqrt{\frac{1}{6}}\tilde{F}. 
\end{gather}
In Eq.~(\ref{eqn:fullamplitude}), the term with $\bar\Sigma$ introduces the sum over $\bar M\bar{B}$ states considered before. Since the poles (consider one channel for simplicity) come from $1-VG=0$, and are the same for particle and antiparticles, the combination $VG$, and hence $tG$, entering Eq.~(\ref{eqn:fullamplitude}), is the same for particle and antiparticles, and then, $T_{\pi\Sigma}=T_{\pi\bar{\Sigma}}$.

In addition to the above contributions [Fig.~\ref{fig:decaymodel}(a, b, c)], we will discuss the effect of $\Sigma\bar\Sigma$ coupling to $p\bar p$, depicted in the Fig.~\ref{fig:decaymodel}(d). This is because the $p\bar p$ has an enhancement close to the threshold that is attributed to the resonance $X(1835)$, which is seen in the decays of $J/\psi\to\gamma\pi^+\pi^-\eta'$ ~\cite{Ablikim:2010au,Alexander:2010vd} and $J/\psi\to p\bar p\gamma$~\cite{BESIII:2011aa}. For the latter decay, they see an enhancement in the $p\bar p$ mass distribution close to threshold. The $\Sigma\bar\Sigma$ will couple to $p\bar{p}$ in coupled channels. So, any pole in the $p\bar{p}\to p\bar{p}$ will also be present in the $p\bar{p}\to \Sigma\bar{\Sigma}$ amplitude. By taking into account the $\Sigma\bar{\Sigma}$ coupling to $p\bar{p}$, Eq.~(\ref{eqn:fullamplitude}) can be rewritten as,
\begin{equation}
\mathcal{M}(M_{\pi\Sigma},M_{\pi\bar\Sigma})=V_p\left( h_{\pi\Sigma} + T_{\pi\Sigma} + T_{\pi\bar\Sigma} + T_{p\bar{p}} \right) ,\label{eqn:fullamplitudenew}
\end{equation}
where,
\begin{equation}
T_{p\bar{p}}=\frac{a}{M_{\Sigma\bar{\Sigma}}-M_{X}+i\frac{\Gamma_{X}}{2}}
\end{equation}
where the $M_X=1835$ MeV and $\Gamma_X=100$ MeV are the mass and width of the resonance $X(1835)$~\cite{pdg}, and the normalization $a$ stands for the amplitude strength. This should not disturb much our result, because the $\Sigma\bar{\Sigma}$ invariance mass $M_{\Sigma\bar{\Sigma}}>2M_\Sigma=2393$ MeV, 560 MeV larger than the mass of $X(1835)$. This is very far and should not have any effect. Yet, we are going to show that even in an extreme case this will not have any effect on the $\pi\Sigma$ mass distribution.  

\subsection{The final state interaction}
Based on the chiral Lagrangian for meson-baryon interactions and the $N/D$ method, the full set of transition  matrix elements with the coupled channels in $I=1$, $\bar{K}N$, $\pi\Sigma$, $\pi\Lambda$, $\eta\Sigma$ and $K\Xi$, can be expressed by means of the on shell factorized Bethe Salpeter (BS) equation,
\begin{equation}
t=\left[1-VG\right]^{-1}V, \label{eq:tm}
\end{equation} 
where the matrix $V$ is obtained from the lowest order meson baryon chiral Lagrangian~\cite{Gasser:1983yg,Bernard:1995dp},
\begin{equation}
V_{ij}(I=1)=-F_{ij}\frac{1}{4f^2}(k^0+{k'}^0), \label{eq:vij}
\end{equation}
where the magnitudes $k^0$ and  ${k'}^0$ are the initial and final energies of the mesons,  
and the symmetrical coefficients $F_{ij}$ are shown in Table~5 of Ref.~\cite{Oset:1997it}. The value $f=1.15f_\pi$ with $f_\pi=93$ MeV, common to all channels, was used in Ref.~\cite{Oset:1997it}, leading to a good fit to the data. 
The loop function $G$ stands for a diagonal matrix with elements: 
\begin{eqnarray}
 \label{Loop_integral}
G_l&=&{\rm i}\int \frac{d^4q}{{(2\pi)}^4}\frac{M_l}{E_l(q)}\frac{1}{k^0+p^0-q^0-E_l(q)+{\rm i}\epsilon} \nonumber \\
&&\times\frac{1}{q^2-m_l^2+{\rm i}\epsilon} \nonumber\\
&=&\int \frac{d^3q}{{(2\pi)}^3}\frac{M_l}{2\omega_l(q)E_l(q)}\frac{1}{k^0+p^0-q^0-E_l(q)+{\rm i}\epsilon}  \nonumber \\
\end{eqnarray}
where $M_l$ and $m_l$ are the baryon and meson masses of the $``l"$ channel, and the cut-off $|\vec{q}_{\rm max}|=630$ MeV is used as in Ref.~\cite{Oset:1997it}.

Finally, the invariant mass distribution $\chi_{c0}(1P)\to \bar\Sigma\,\Sigma\,\pi$ reads
\begin{align}\label{eqn:dGammadM}
\frac{d^2\Gamma_j}{dM^2_{\pi\Sigma}dM^2_{\pi\bar\Sigma}}
=\frac{1}{(2\pi)^3}\frac{4M^2_\Sigma}{32M^3_{\chi_{c0}}} \left|\mathcal{M}(M_{\pi\Sigma}, M_{\pi\bar\Sigma})\right|^2\,,
\end{align}
where $M_{\pi\Sigma}$ and $M_{\pi\bar\Sigma}$ are the invariant mass of $\pi\Sigma$ and $\pi\bar\Sigma$. For a given value of $M^2_{\pi\Sigma}$, the range of $M^2_{\pi\bar\Sigma}$ is defined as,
\begin{eqnarray}
(M^2_{\pi\bar\Sigma})_{\rm max}\!&=&\! \left(E^*_\pi+E^*_{\bar\Sigma}\right)^2-\left(\sqrt{E^{*2}_\pi-m^2_\pi}-\sqrt{E^{*2}_{\bar\Sigma}-m^2_{\bar\Sigma}}\right)^2,  \nonumber \\ 
(M^2_{\pi\bar\Sigma})_{\rm min}\! &=&\! \left(E^*_\pi+E^*_{\bar\Sigma}\right)^2-\left(\sqrt{E^{*2}_\pi-m^2_\pi}+\sqrt{E^{*2}_{\bar\Sigma}-m^2_{\bar\Sigma}}\right)^2,  \nonumber \\ 
\end{eqnarray} 
here $E^{*}_\pi=(M^2_{\pi\Sigma}-M^2_\Sigma+m^2_\pi)/2M_{\pi\Sigma}$ and $E^{*}_{\bar\Sigma}=(M^2_{\chi_{c0}}-M^2_{\pi\Sigma}-M^2_{\bar\Sigma})/2M_{\pi\Sigma}$ are the energies of $\pi$ and $\bar\Sigma$ in the $M_{\pi\Sigma}$ rest frame.

The invariance mass of $\Sigma\bar{\Sigma}$ can be related to the $M_{\pi\Sigma}$ and $M_{\pi\bar\Sigma}$ by
\begin{equation}
M^2_{\Sigma\bar\Sigma}=M^2_{\chi_{c0}}+m^2_\pi+2M^2_\Sigma-M^2_{\pi\bar\Sigma}-M^2_{\pi\Sigma}.
\end{equation}

The on shell factorized BS equation of Eqs.~(\ref{eq:tm}, \ref{eq:vij}) can be obtained from the Chew-Mandelstam $N/D$ method \cite{Chew:1960iv} by neglecting the left hand cut (which is normally included in the factor $N$), but looking explicitly at the unitarity cut which is included in $D$, and is calculated using a dispersion relation. This is done explicitly in Refs.~\citep{ollerulf, nsd}. In Ref.~\cite{ollerulf}, the influence of the left hand cut in this interaction is found very small. But, even in case where this is not necessarily true, the distance of the left hand cut to the physical energies renders its contribution in the dispersion relation rather energy independent such that it can be accommodated by means of a suitable choice of subtraction constants in the dispersion relation, which are adjusted to data. The effect of the left hand cut can also be addressed within the BS equation as discussed in Refs.~\citep{ollerulf, nsd,Lacour:2009ej}, or using the inverse amplitude method \cite{Dobado:1996ps, Pelaez:2015qba}. In Eq.~(\ref{eq:vij}), the kernel used for the BS equation comes from the lowest order chiral Lagrangian \cite{nsd}. There has been much recent work including in the kernel the term from higher order chiral Lagrangian \cite{Borasoy:2005ie,Oller:2006jw,
Borasoy:2006sr,hyodonew,kanchan,hyodorev,ollerguo,maimeissner,feijoo}. However, as shown in Refs.~\cite{Borasoy:2006sr,hyodorev}, the effect of higher orders in this interaction is very small, and can also be easily accommodated by suitable changes in the subtraction constants (or equivalently cut off) in the dispersion integral leading to the G function (see also Ref.~\citep{hyodoHadron} for a recent review on this interaction and the two states of the $\Lambda(1405)$). Altogether, as shown in Ref.~\cite{Oset:1997it}, by using the lowest order kernel of Eq.~(\ref{eq:vij}) and a suitable cut off to regularize the loops (G function), an excellent description of the low energy $\bar{K}N$ data and cross sections to coupled channels, in a wider range than the one investigated here, was obtained.

\section{Results and Discussion}\label{sec:results}
\begin{figure}[!htb]
\centering
  \includegraphics[width=0.45\textwidth]{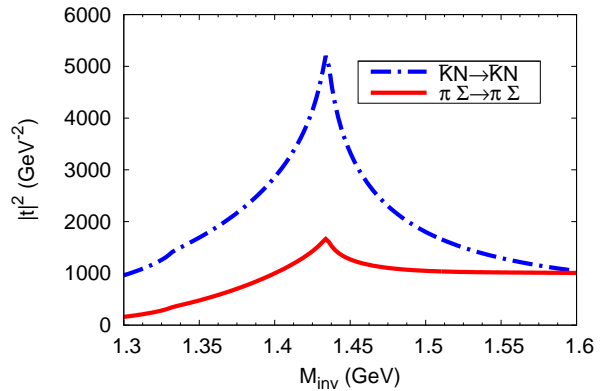}
\caption{(Color online) Module squared of the $t_{\bar{K}N, \bar{K}N}$ and $t_{\pi\Sigma,\pi\Sigma}$ as a function of the invariant mass of $\bar{K}N$ (or $\pi\Sigma$) system. } 
  \label{fig:moduleT}
\end{figure}

\begin{figure}[!htb]
\centering
  \includegraphics[width=0.45\textwidth]{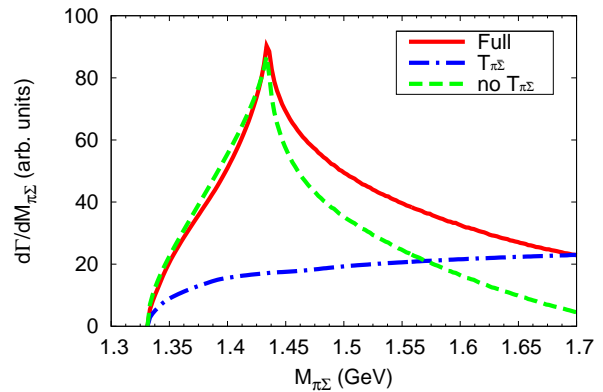}
\caption{(Color online) The $\pi\Sigma$ invariant mass distributions for the $\chi_{c0}(1P)\to\bar\Sigma\Sigma\pi$ decay with $R=1$.} 
  \label{fig:dwidth}
\end{figure}

In this section, we present our results for the process $\chi_{c0}\rightarrow\bar \Sigma\Sigma\pi$. First, we show the module squared of the amplitudes $|t_{\bar{K}N, \bar{K}N}|^2$ and $|t_{\pi\Sigma,\pi\Sigma}|^2$ in $I=1$ in Fig.~\ref{fig:moduleT}. The cusp aspects are found at the $\bar{K}N$ threshold, the same as the result of Ref.~\cite{luispho2}.

Next, we predict the $\pi\Sigma$ invariant mass distribution for the $\chi_{c0}(1P)\to\bar\Sigma\Sigma\pi$ decay in Fig.~\ref{fig:dwidth}.
We have two parameters $\tilde{D}$ and $\tilde{F}$. 
Since we have an arbitrary normalization, we can work with $R=\tilde{F}/\tilde{D}$,
 and include the weight of $\tilde{D}$ into the $V_p$ factor. Hence, up to an arbitrary normalization, our results depend on the ratio $R$. The idea is to evaluate the mass distributions for different values of this ratio, and see if the strong cusp structure remains.
In Fig.~\ref{fig:dwidth}, the red solid line stands for the result of our full model, and the blue dashed-dotted line shows the contribution of the $\pi\bar\Sigma$ interaction [the term $T_{\pi\bar\Sigma}$ of the Eq.~(\ref{eqn:fullamplitude})]. Finally, the green dashed line corresponds to the contribution of the tree level and $\pi\Sigma$ interaction [the term $h_{\pi\Sigma}+T_{\pi\Sigma}$ of the Eq.~(\ref{eqn:fullamplitude})]. Here, we take $R=1$. 
We observe a strong cusp structure around the $\bar{K}N$ threshold when the $\pi\Sigma$ interaction is taken into account. We can see that considering the interaction of the $\pi\bar{\Sigma}$ in addition does not practically influence the structure seen when one considers the $\pi\Sigma$ interaction alone. This is because when we choose the invariant mass of $\pi\Sigma$ around the peak in the figure, the $\pi\bar{\Sigma}$ invariant mass is very different and is not affected by this structure around the $\bar{K}N$ threshold. 

\begin{figure}[!htb]
\centering
  \includegraphics[width=0.45\textwidth]{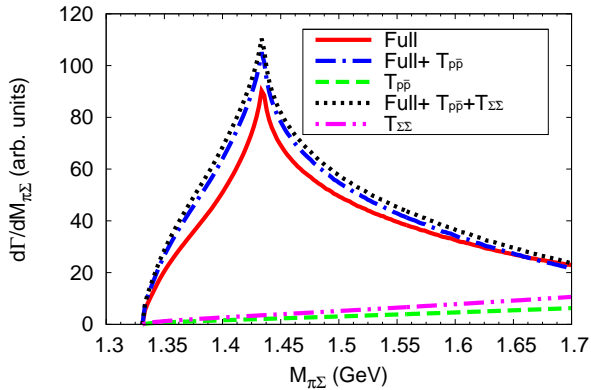}
\caption{(Color online) The $\pi\Sigma$ invariant mass distributions for the $\chi_{c0}(1P)\to\bar\Sigma\Sigma\pi$ decay with $R=1$ by taking into account the effects of $\Sigma\bar{\Sigma}$ coupling to  $p\bar{p}$, and of the $\Sigma\bar{\Sigma}$ interaction.} 
  \label{fig:dwidth_new}
\end{figure}
We also show the effect of $\Sigma\bar{\Sigma}$ coupling to  $p\bar{p}$ on the $\pi\Sigma$ mass distribution in Fig.~\ref{fig:dwidth_new}. The red solid line stands for the result of our full model, the blue dashed-dotted line corresponds to the result by adding the contribution of $\Sigma\bar{\Sigma}$ coupling to  $p\bar{p}$, and green dashed line shows the contribution of $\Sigma\bar{\Sigma}$ coupling to $p\bar{p}$ alone. The value of normalization has been chosen $a=300$ MeV such that the effect of this term in the invariant mass distribution is sizeable, of the order of $25\%$ increase in the mass distribution, in spite of the large mass difference between $\Sigma\bar{\Sigma}$ and $X(1835)$ commented above.
As we can see, the shape of the $\pi\Sigma$ mass distribution does not change, still showing a clear cusp around $\bar{K}N$ threshold. Thus, we will neglect the effect of $\Sigma\bar{\Sigma}$ coupling to  $p\bar{p}$ in the following.   
\begin{figure}[!htb]
\centering
  \includegraphics[width=0.45\textwidth]{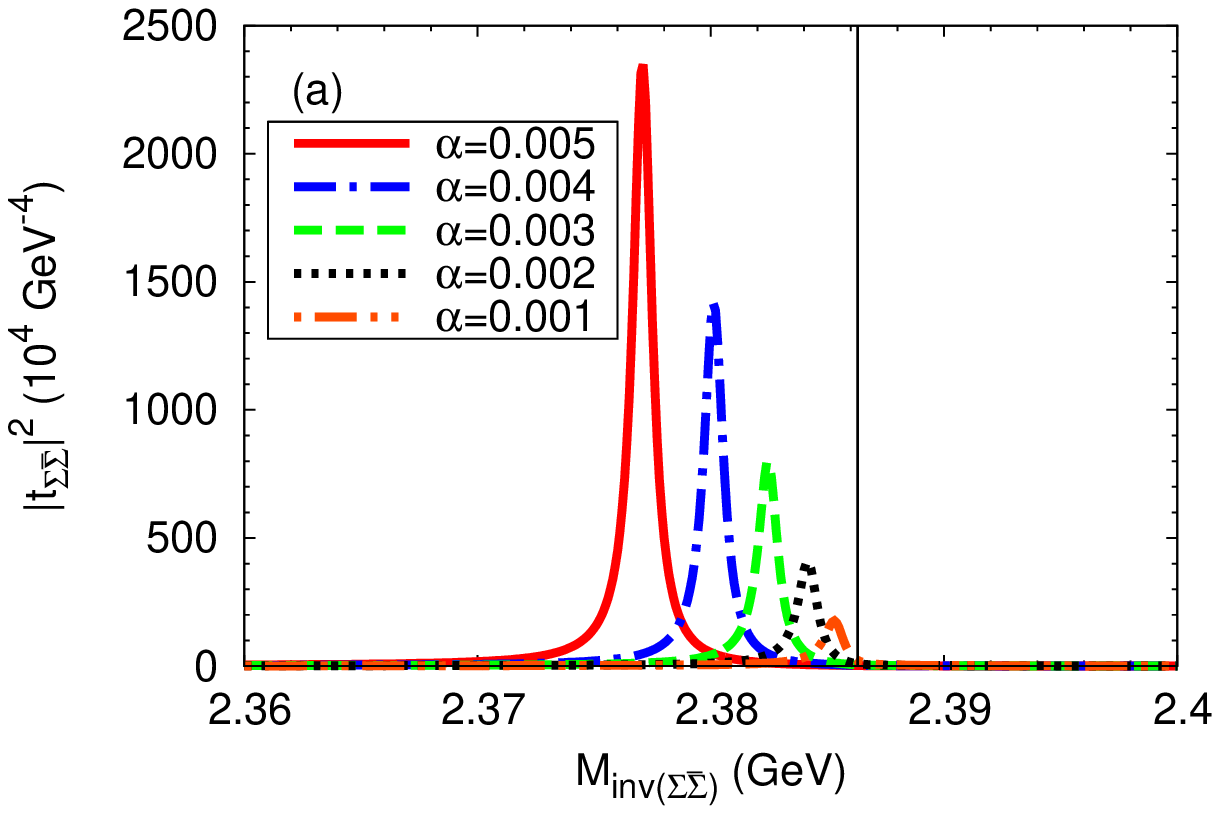}
  \includegraphics[width=0.45\textwidth]{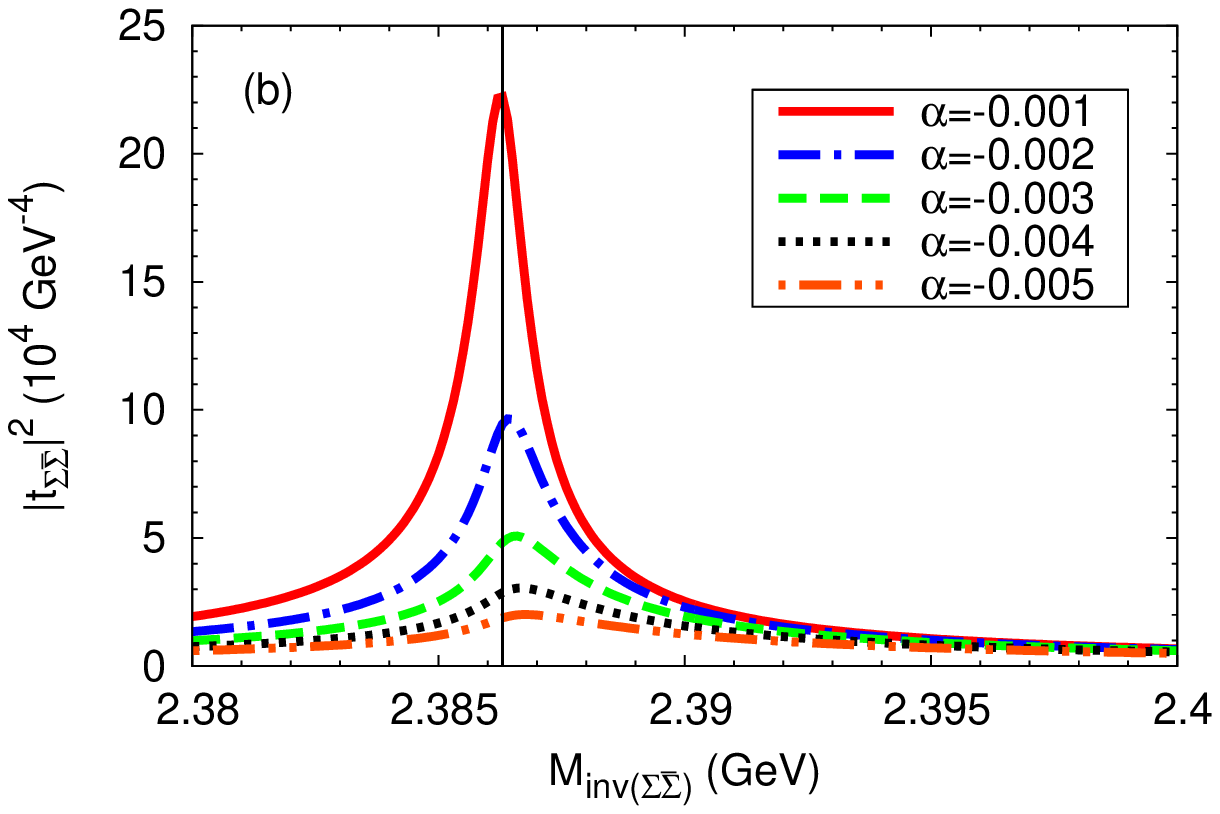}
\vspace{0.5cm}
\caption{(Color online) Module squared of the $t_{\Sigma\bar{\Sigma}}$ for different values of $\alpha$: (a) $\alpha>0$, (b) $\alpha<0$. The black vertical lines stand for the threshold of $\Sigma\bar{\Sigma}$.} 
  \label{fig:moduleT_SSb}
\end{figure}

There is another test that we can conduct. It could happen that the $\Sigma\bar{\Sigma}$ interaction has some sharp structure at threshold. This could be as a consequence of an attractive $\Sigma\bar{\Sigma}$ interaction which barely binds the system or fails shortly to do it. In the first case, we would get a bound state, in the second one a strong cusp structure on the $\Sigma\bar{\Sigma}$ amplitude. We consider this interaction by taking again Fig.~\ref{fig:decaymodel}(d), and for the $\Sigma\bar{\Sigma}$ scattering matrix we take the amplitude that stems from a potential $V_{\Sigma\bar{\Sigma}}$,
\begin{equation}
t_{\Sigma\bar{\Sigma}}=\frac{1}{V^{-1}_{\Sigma\bar{\Sigma}}-G_{\Sigma\bar{\Sigma}}(M_{\Sigma\bar{\Sigma}})}, \label{eq:tSSb}
\end{equation} 
where $G_{\Sigma\bar{\Sigma}}$ is now given by,
\begin{equation}
G_{\Sigma\bar{\Sigma}}(M_{\Sigma\bar{\Sigma}})=\int \frac{d^3q}{(2\pi)^3}\frac{M^2_\Sigma}{E^2(q)}\frac{1}{M_{\Sigma\bar{\Sigma}}-2E(q)+i\epsilon},\label{eq:gSSb}
\end{equation}
which we regularize by a typical cut off $|\vec{q}_{\rm max}|=600$ MeV~\cite{Oset:1997it} (changes in $|\vec{q}_{\rm max}|$ can be reabsorbed by changes in $V_{\Sigma\bar{\Sigma}}$).

A pole at threshold requires $V^{-1}_{\Sigma\bar{\Sigma}}=G_{\Sigma\bar{\Sigma}}(2M_\Sigma)$. Then we take,
\begin{equation}
V^{-1}_{\Sigma\bar{\Sigma}}=G_{\Sigma\bar{\Sigma}}(2M_\Sigma)+\alpha M^2_\Sigma,
\end{equation}
and for $\alpha>0$ we have a bound state, while for $\alpha<0$ we get the cusp structure. 
In Figs.~\ref{fig:moduleT_SSb}(a) and (b), we plot $|t_{\Sigma\bar{\Sigma}}|^2$ to show the structure that we have created. Indeed, a strong cusp structure around the $\Sigma\bar{\Sigma}$ threshold is observed for negative $\alpha$. We take a value $\alpha=-0.001$, which leads to a very pronounced cusp structure, to do the following exercise, but the same conclusions are obtained for any value of $\alpha$. Then to take into account the contribution of the new structure in the $\pi\Sigma$ mass distribution, we add to Eq.~(\ref{eqn:fullamplitudenew}) the term,
\begin{equation}
T_{\Sigma\bar{\Sigma}}=h_{\pi\Sigma}G_{\Sigma\bar{\Sigma}}(M_{\Sigma\bar{\Sigma}})t_{\Sigma\bar{\Sigma}}(M_{\Sigma\bar{\Sigma}}), \label{eq:ampSSb}
\end{equation}
and Eq.~(\ref{eqn:fullamplitudenew}) becomes now,
\begin{equation}
\mathcal{M}(M_{\pi\Sigma},M_{\pi\bar\Sigma})=V_p\left( h_{\pi\Sigma} + T_{\pi\Sigma} + T_{\pi\bar\Sigma} + T_{p\bar{p}} + T_{\Sigma\bar{\Sigma}} \right) .\label{eqn:fullamplitudenewnew}
\end{equation}
We plot in Fig.~\ref{fig:dwidth_new} the result of adding this new structure, which is shown by the black dotted line. The magenta dashed-dotted-dotted line stands for the contribution of this $\Sigma\bar{\Sigma}$ interaction alone.
As we can see, there is a small effect in the $\pi\Sigma$ mass distribution, but what is more important, the $\pi\Sigma$ cusp structure has not been spoiled. Since the $\Sigma\bar{\Sigma}$ will annihilate, the $V_{\Sigma\bar{\Sigma}}$ potential should also contain an imaginary part. For values of Im$V_{\Sigma\bar{\Sigma}}$ of the order of Re$V_{\Sigma\bar{\Sigma}}$, the structures in the $\Sigma\bar{\Sigma}$ amplitude are softened and, a fortiori, the cusp structure in the $\pi\Sigma$ invariant mass remains unchanged.

\begin{figure}[!htb]
\centering
  \includegraphics[width=0.45\textwidth]{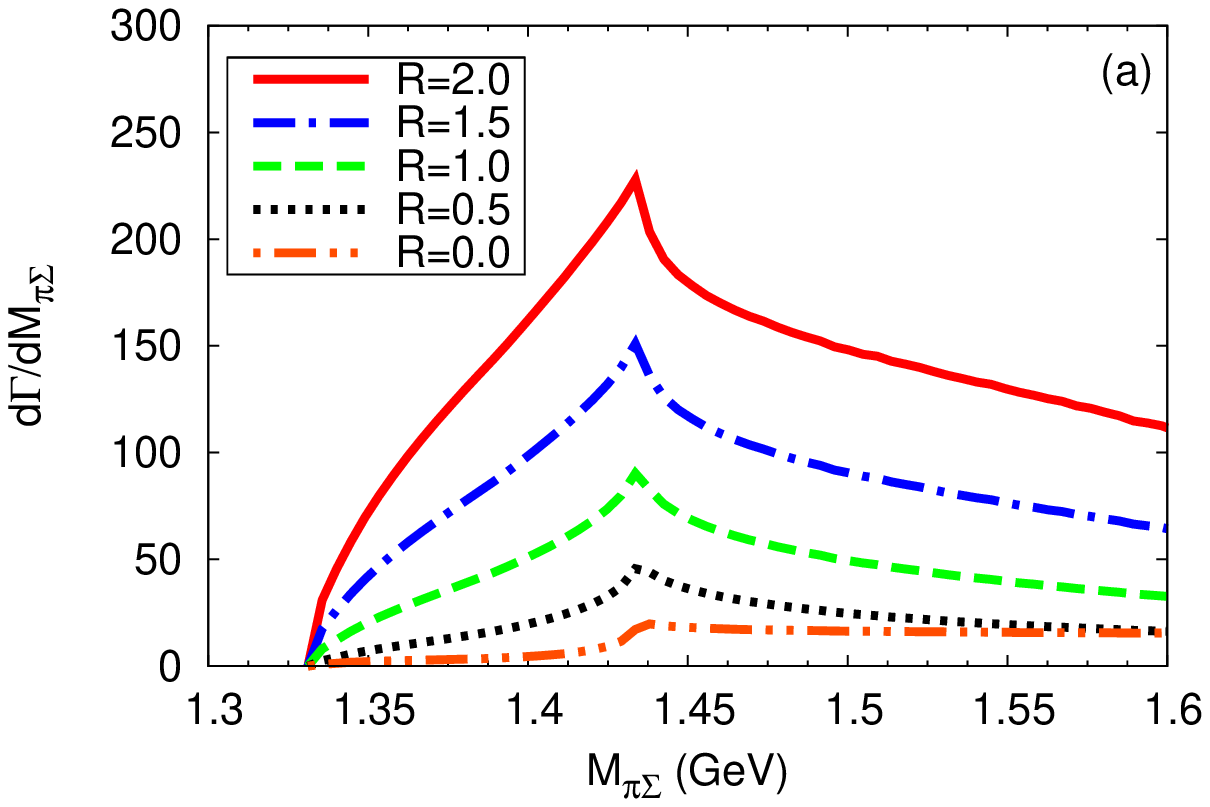}
  \includegraphics[width=0.45\textwidth]{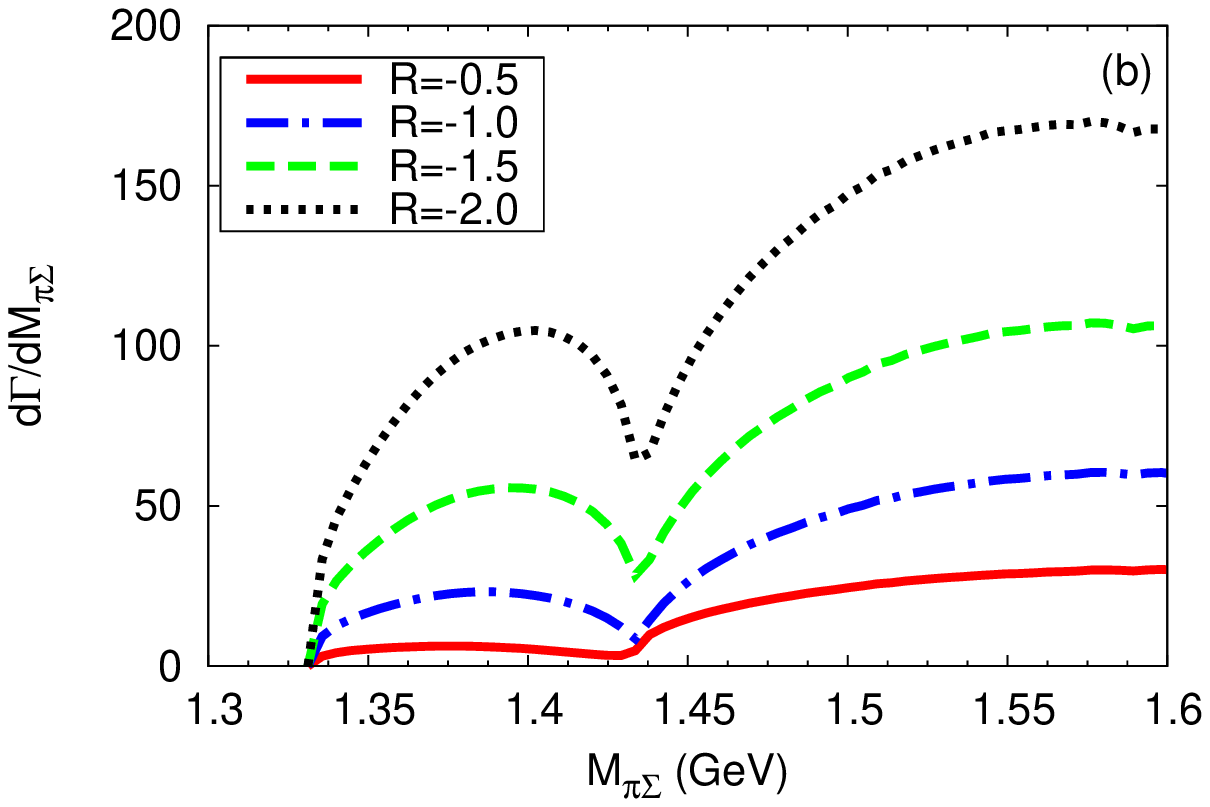}
\caption{(Color online) The $\pi\Sigma$ invariant mass distributions for the $\chi_{c0}(1P)\to\bar\Sigma\Sigma\pi$ decay by including the full contributions with different values of $R$. (a): $R$ is positive; (b) $R$ is negative.} 
  \label{fig:depend}
\end{figure}

As we do not know the exact value of the ratio $R$, we calculate the differential decay width of this process with different values of $R$, and this is depicted in Fig.~\ref{fig:depend}. 
We can see that for a wide range of values of $R$, the strong cusp structure around the $\bar{K}N$ threshold remains. It is interesting to observe that for positive values of $R$, we have a peak, but for negative values of $R$, the peak is inverted, becoming a sharp dip. One must trace that to the isoscalar coefficients in Table~\ref{tab:weight}, If one takes $R$ negative, then the $t_{\pi\Sigma,\pi\Sigma}$ amplitude appearing in the $T_{\pi\Sigma}$ of Eq.~(\ref{eqn:fullamplitude}) gets multiplied by  $h_{\pi\Sigma}$, which is now negative, while the factor is positive when $R$ is positive.

\begin{figure}[!htb]
\centering
  \includegraphics[width=0.45\textwidth]{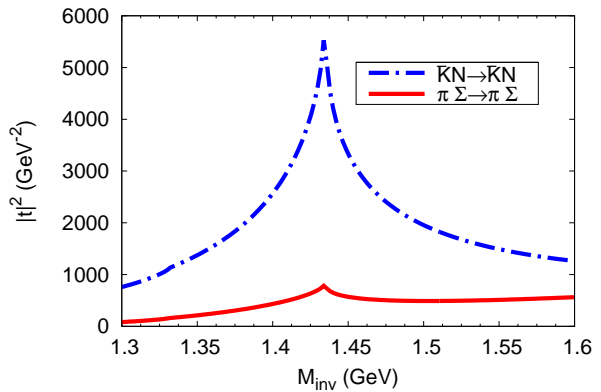}
\caption{(Color online) Same as Fig.~\ref{fig:moduleT} but with the coefficient $F_{ij}$ in Eq.~(5) of Ref.~\cite{luispho2} and dimensional regularization $G$ for the channels of $\bar{K}N$, $\pi\Sigma$, $\pi\Lambda$.} 
  \label{fig:moduleT_roca}
\end{figure}

\begin{figure}[!htb]
\centering
  \includegraphics[width=0.45\textwidth]{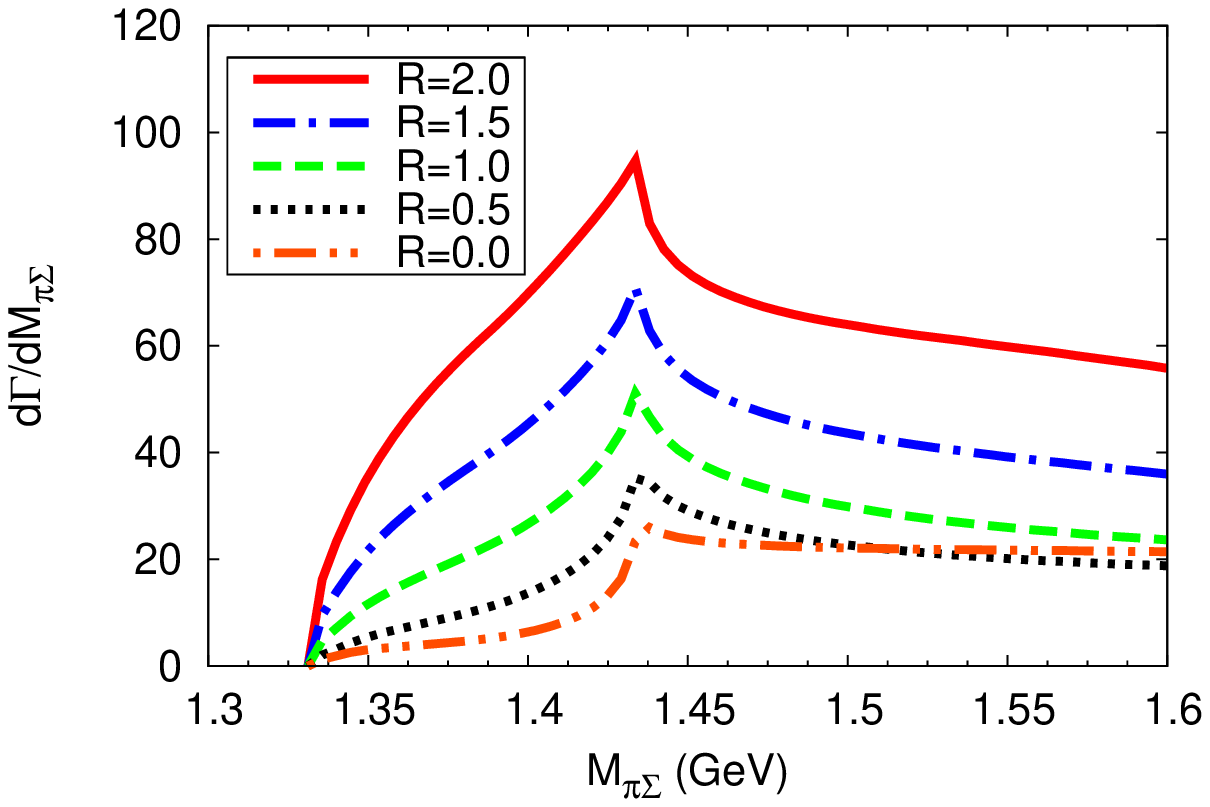}
\caption{(Color online) Same as Fig.~\ref{fig:depend}(a) but with the coefficient $F_{ij}$ in Eq.~(5) of Ref.~\cite{luispho2} and dimensional regularization $G$ for the channels of $\bar{K}N$, $\pi\Sigma$, $\pi\Lambda$.} 
  \label{fig:depend_roca}
\end{figure}

We also show the results by using the coefficient $F_{ij}$ given in Eq.~(5) of Ref.~\cite{luispho2}. The loop function $G$ in Eq.~(\ref{eq:tm}) is obtained with the dimensional regularization and the subtraction constants $a_\mu$ was also taken from this reference~\footnote{In Ref.~\cite{luispho2}, only three channels of $\bar{K}N$, $\pi\Sigma$, $\pi\Lambda$ are considered, so we use the new coefficients  $F_{ij}$ and dimensional regularization $G$ for those three channels, and keep the same for the other two channels.}. First, we re-plot Fig.~\ref{fig:moduleT} with the new input, which is shown in Fig.~\ref{fig:moduleT_roca}. Both the shape of the $\bar{K}N\to\bar{K}N$ and $\pi\Sigma\to\pi\Sigma$ amplitude module squared are same. 
The strength of the $\bar{K}N\to\bar{K}N$ amplitude is not much affected, but
the one of $\pi\Sigma\to\pi\Sigma$ is reduced by about a factor of two 
if the coefficients and the dimensional regularization are used. With the new input, we present the differential decay width of this process for positive $R$ in Fig.~\ref{fig:depend_roca}. From this figure, we can see that the shapes and the cusp position are same as in Fig.~\ref{fig:depend}.

One may wonder why not to make the reaction from $J/\psi$ decay, since the SU(3) symmetry would be the same. The reason is that the $\chi_{c0}$ has quantum number $J^P=0^+$. Then it decays into $\bar{\Sigma}$ ($1/2^-$) (the negative parity because it is antiparticle), $\Sigma$ ($1/2^+$) and $\pi$ ($0^-$). Then the decay can be accommodated with $L=0$. If we start with the $J/\psi$ ($1^-$), we need $L=1$  to restore the parity and one has a more complicated structure to couple spins and angular momenta. In principle, $L=0$ should be also favored with respect to $L=1$, and this could explain why the width of $\chi_{c0}$ to $p\bar{p}\pi$ is bigger than the one of $\chi_{c0}$ to $p\bar{p}$, while in the case of $J/\psi$, the rate of $p\bar{p}$ decay is bigger than that of $p\bar{p}\pi$~\cite{pdg}.


There is another aspect that one might like to bring to discussion and this is if the strong cusp can be associated to a resonance. Technically, one does not have a pole in the second Riemann sheet, but from the theoretical point of view, a state with small binding  and one barely unbound, reflecting in a strong cusp, are obtained with small changes in the parameters of the theory and reflect the same physics. It is a question of criterion to adopt a classification for such a state. The fact is that the situation is identical to the one found for the $a_0(980)$ resonance, that both in the theory~\citep{ramonet,  dany,Pelaez:2006nj,Pelaez:2003dy,Baru:2003qq,Branz:2008ha,Wolkanowski:2015lsa, luispho2} and in experiment~\cite{PRL93} shows a strong cusp structure around the $K\bar{K}$ threshold, and is classified as a standard resonance.

\section{Conclusions}\label{sec:conclusions}
In this paper, we have suggested to use the $\chi_{c0}(1P)\to\bar\Sigma\Sigma\pi$ reaction as a test of the existence of an $I=1$, $S=-1$, and $J^P=1/2^-$ resonance close to the $\bar{K}N$ threshold. The state appears in all theoretical works using the chiral unitary approach, but it is border line, meaning that in some works it appears as a weakly bound state, while in others, as a lightly unbound or virtual state, but in all cases, it is reflected as a strong cusp around the $\bar{K}N$ threshold.

The reaction chosen guarantees that the $\pi\Sigma$ state is produced in $I=1$, and hence it is a filter of isospin, facilitating the observation of states in this sector. We have shown that, up to an arbitrary normalization, the results depend on the ratio of $\tilde{F}/\tilde{D}$ which we do not know. But we observed that in a large range of values of this ratio, the cusp structure is always observed, as a peak, when $\tilde{F}/\tilde{D}$ is positive, or as a sharp dip when $\tilde{F}/\tilde{D}$ is negative. We have also shown that the values estimated are well within the range of possible measurements of BESIII, and the implementation of the experiment would be most welcome.

\section*{Acknowledgments}
One of us, E. O., wishes to acknowledge support from
the Chinese Academy of Science in the Program of Visiting Professorship for Senior International Scientists (Grand No. 2013T2J0012). This work is partly supported
by the National Natural Science Foundation of China under Grant Nos. 11505158, 11475227, and
the Spanish Ministerio de Economia y Competitividad
and European FEDER funds under the contract number
FIS2011-28853-C02-01 and FIS2011-28853-C02-02, and
the Generalitat Valenciana in the program Prometeo II,
2014/068. This work is also supported by the
Open Project Program of State Key Laboratory of Theoretical Physics,
Institute of Theoretical Physics, Chinese Academy of Sciences, China
(No.Y5KF151CJ1), and the China Postdoctoral Science Foundation (No. 2015M582197). 
We acknowledge the support of the European
Community-Research Infrastructure Integrating Activity
Study of Strongly Interacting Matter (acronym HadronPhysics3, Grant Agreement n. 283286) under 
the Seventh Framework Programme of EU.

\end{document}